\documentclass[12pt]{article}
\usepackage{graphicx} 
\usepackage{amsmath,amssymb}
\usepackage{hyperref}
\usepackage{cite}
\usepackage{geometry}

\title{"Quantum-Enhanced Conformal Methods for Multi-Output Uncertainty: A Holistic Exploration and Experimental Analysis"}
\author{
\textbf{Davut Emre TASAR}\\
{\small \textit{University of Navarra, Applied Engineering PhD}}\\
{\small Pamplona, Spain}\\
{\small \texttt{dtasar@alumni.unav.es}}\\
{\small ORCID: 0000-0002-7788-0478}
}

\date{}

\begin{document}

\maketitle

\section*{Abstract}

\vspace{0.5em}
\noindent
\textit{Quantum computing} introduces unique forms of randomness arising from measurement processes, gate noise, and hardware imperfections. Ensuring reliable \textit{uncertainty quantification} in such quantum-driven or quantum-derived predictions is an emerging challenge. In classical machine learning, \textit{conformal prediction} has proven to be a robust framework for distribution-free uncertainty calibration, often focusing on univariate or low-dimensional outputs. Recent advances (e.g., \cite{ParkSimeone2023, Xu2024, Feldman2021}) have extended conformal methods to handle multi-output or multi-dimensional responses, addressing sophisticated tasks such as time-series, image classification sets, and quantum-generated probability distributions. However, bridging the gap between these powerful conformal frameworks and the high-dimensional, noise-prone distributions typical of quantum measurement scenarios remains largely open.

In this paper, we propose a unified approach to harness \textit{quantum conformal methods} for multi-output distributions, with a particular emphasis on two experimental paradigms: 
(\textit{i}) a standard 2-qubit circuit scenario producing a four-dimensional outcome distribution, and 
(\textit{ii}) a multi-basis measurement setting that concatenates measurement probabilities in different bases (Z, X, Y) into a twelve-dimensional output space. 
By combining a multi-output regression model (e.g., random forests) with \textit{distributional conformal prediction}, we validate coverage and interval-set sizes on both simulated quantum data and multi-basis measurement data. Our results confirm that classical conformal prediction can effectively provide coverage guarantees even when the target probabilities derive from inherently quantum processes. 
Such synergy opens the door to next-generation quantum-classical hybrid frameworks, providing both improved interpretability and rigorous coverage for quantum machine learning tasks. \textbf{All codes and full reproducible Colab notebooks} are made available at
\href{https://github.com/detasar/QECMMOU}{https://github.com/detasar/QECMMOU}.

\vspace{0.5em}
\noindent
\textbf{Keywords:} Quantum Computing, Conformal Prediction, Multi-Output Regression, Distribution-Free Coverage, Multi-Basis Measurement, Quantum Machine Learning.

\clearpage

\section{Introduction}

\vspace{0.5em}
\noindent
\textbf{Motivation and Background.}\quad
Quantum computing leverages superposition and entanglement of qubits to (potentially) solve problems at scales intractable for classical machines. However, even as Noisy Intermediate-Scale Quantum (NISQ) devices progress \cite{Huang2023Learning}, the inherent measurement randomness, gate errors, and hardware noise complicate the generation of stable output probabilities. Classical \textit{conformal prediction} (CP) has emerged as a powerful, distribution-free framework that can quantify predictive uncertainty in a statistically rigorous way \cite{Cherubin2020, Angelopoulos2024, Bai2022GenFunc}. 
Until recently, most conformal approaches addressed single-dimensional outputs or classification tasks \cite{Bethell2023}. 
Yet, the need to provide set-valued predictions or region-based coverage for multi-output data is rapidly growing \cite{Carlsson2022, Xu2024}, especially in quantum contexts where the natural outputs (e.g., measurement distributions) are inherently multi-dimensional. 

\vspace{0.5em}
\noindent
\textbf{Quantum-Specific Challenges.}\quad
Predicting the outcome distribution of a quantum circuit entails dealing with:
\begin{itemize}
    \item \textit{Stochasticity of measurement.} A 2-qubit circuit, for instance, yields a random distribution over $\{00,01,10,11\}$. Each execution (shot) collapses the state, introducing inherent randomness.
    \item \textit{Hardware noise and drifts.} Real quantum devices exhibit gate infidelities and drift over time, causing correlated noise across measurement shots \cite{ParkSimeone2023}.
    \item \textit{High-dimensional expansions.} If measurements are taken in multiple bases (e.g., X, Y, Z), the resulting multi-head distribution can reach dimension $4m$ for $m$ different bases in just a 2-qubit system. 
\end{itemize}
Therefore, guaranteeing a coverage statement like “with $90\%$ probability, the true measurement distribution lies within the predicted region” is non-trivial. 
Recent works \cite{ParkSimeone2023} discuss \textit{probabilistic conformal prediction (PCP)} for quantum models, but typically focus on single-basis or single-dimensional scenarios.

\vspace{0.5em}
\noindent
\textbf{Prior Art in Multi-Output Conformal.}\quad
Meanwhile, multi-output conformal methods in classical machine learning have flourished. 
For instance, ellipsoidal sets for multi-dimensional time series \cite{Xu2024}, multi-output regression intervals \cite{Feldman2021, Carlsson2022}, and adaptive or differentiable conformal solutions \cite{Bai2022GenFunc, Xu2024} represent active directions. 
These techniques ensure finite-sample coverage under minimal assumptions—primarily \textit{exchangeability} of calibration and test points. 
In quantum-like scenarios, exchangeability might hold when each circuit is drawn from the same distribution of gates or the same family of states, so the typical CP framework can be leveraged.

\vspace{0.5em}
\noindent
\textbf{Contributions and Paper Outline.}\quad
In this work, we propose a systematically integrated approach to:
\begin{enumerate}
    \item Generate synthetic data from quantum circuits and multi-basis measurements. One scenario yields a $4$-dim distribution from a 2-qubit measurement (computational basis), while the second scenario concatenates $Z$, $X$, and $Y$ measurement probabilities into a $12$-dim vector.
    \item Apply classical \textit{multi-output regression} to map classical features (circuit-depth, gate counts, etc.) to these quantum-derived probability vectors.
    \item Adopt a \textit{distributional conformal prediction} approach, using a single scalar norm (e.g., $\ell_2$ or $\ell_\infty$) to form coverage sets in $\mathbb{R}^4$ or $\mathbb{R}^{12}$.
    \item Experimentally evaluate how coverage changes with different miscoverage parameters $\alpha$, reporting coverage rates and “volume” (or set size) of these multi-dimensional intervals (hypercubes or hyperspheres).
\end{enumerate}
We show that, under mild assumptions, conformal intervals indeed provide coverage close to $(1-\alpha)$, even though the underlying data stems from quantum measurements. This synergy offers an important step toward quantum-classical hybrid pipelines where classical conformal methods supply robust uncertainty quantification for quantum outputs.

\vspace{0.5em}
\noindent
\textbf{Paper Organization.}\quad
We structure the paper as follows:
\begin{itemize}
    \item \textbf{Section II (Materials and Methods)} details the quantum circuit generation, multi-basis measurement strategies, and the fundamentals of multi-output conformal intervals.
    \item \textbf{Section III (Experiments)} describes how we train random forest regressors on these multi-output distributions, and how we apply distributional conformal thresholds.
    \item \textbf{Section IV (Results and Discussions)} discusses coverage performance, the trade-off with set sizes, and potential limitations with quantum noise.
    \item \textbf{References and Appendices} provide literature context, expansions of conformal proofs, and the relevant code annotations for replicability.
\end{itemize}

\vspace{0.8em}
\noindent
By bridging modern quantum data generation with distribution-free uncertainty quantification, we illuminate a pathway to robust, “error-bounded” quantum machine learning—an essential milestone in advancing practical quantum computing.

\section{Materials and Methods}
\label{sec:materials_and_methods}

In this section, we detail the \emph{methodological} aspects of our study, which bridges quantum circuit data generation with multi-output conformal prediction. Specifically, we describe: 
\begin{enumerate}
    \item The \emph{quantum data generation pipeline} for single-basis (2-qubit) as well as \emph{multi-basis} measurements.
    \item The \emph{classical features} extracted from the quantum circuits.
    \item Our \emph{multi-output regression} strategy.
    \item The \emph{distributional conformal approach} for interval (or region) construction in up to 12 dimensions. 
\end{enumerate}
We also provide a conceptual flowchart in Fig.~\ref{fig:flowchart} to illustrate how these components interconnect. 

\subsection{Overview of the Pipeline}

\begin{figure}[t]
\centering
\includegraphics[width=0.82\textwidth]{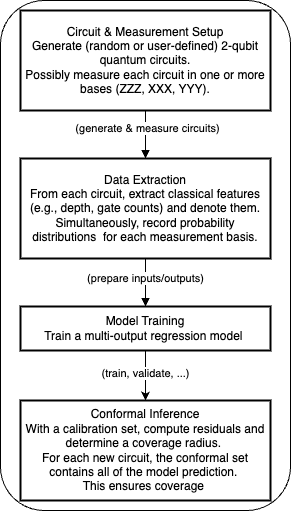} 
\caption{High-level flowchart depicting our method. A random or user-specified 2-qubit quantum circuit is run (possibly in multiple measurement bases), generating a probability vector (4D for single-basis, 12D for triple-basis). We combine gate-level features with these measured vectors to form $(\mathbf{X}, \mathbf{Y})$. A classical regressor is trained and later validated via distributional conformal sets, ensuring coverage near $(1-\alpha)$.}
\label{fig:flowchart}
\end{figure}

Figure~\ref{fig:flowchart} summarizes the overarching pipeline, from quantum circuit generation to final conformal sets. The pipeline is divided into four major phases:
\begin{enumerate}
    \item \textbf{Circuit and Measurement Setup:} We design a random 2-qubit quantum circuit, optionally measuring it in one or more bases (e.g., $\{Z, X, Y\}$).
    \item \textbf{Data Extraction:} We gather classical features (gate counts, depth, etc.) in an $\boldsymbol{X}$ matrix, along with measured distribution vectors (probabilities) in a $\boldsymbol{Y}$ matrix.
    \item \textbf{Model Training:} A multi-output regression model is trained to approximate $f: \boldsymbol{X}\!\to\!\boldsymbol{Y}$.
    \item \textbf{Conformal Inference:} Using a calibration subset, we derive coverage thresholds (radii) for distributional conformal sets in $\mathbb{R}^4$ or $\mathbb{R}^{12}$. We then test coverage on new circuits.
\end{enumerate}

\subsection{Quantum Circuit Generation}
\label{sec:quantum_circuit_generation}

\paragraph{2-Qubit Architecture.} 
We focus on 2-qubit circuits due to their simplicity and ability to demonstrate non-trivial entanglement. Each qubit is initialized to $\lvert 0\rangle$. The circuit depth $D$ (ranging from 1 to 8 in some experiments) dictates the number of gate layers. A depth-1 circuit could apply one or two gates (including possible controlled gates), whereas a depth-8 circuit can have up to 16 or more operations, depending on the random draws.

\paragraph{Random Gate Selection.} 
Following \cite{ParkSimeone2023} and others, we employ a pseudorandom scheme (e.g., Qiskit's \texttt{random\_circuit()}) that draws from common gates (\texttt{H, X, Y, Z, RX, RZ, CX, \dots}). This ensures a diverse distribution of unitary transformations. At the end of each circuit, we apply a measurement operation:
\begin{itemize}
    \item \emph{Single-Basis (Z) scenario:} We measure in the computational (Z) basis, obtaining probabilities for $\{\lvert 00\rangle,\lvert 01\rangle,\lvert 10\rangle,\lvert 11\rangle\}$.
    \item \emph{Multi-Basis scenario:} We replicate the circuit or apply basis transformations to measure in $X$, $Y$, or $Z$ bases. For instance, measuring in the $X$ basis typically involves a Hadamard on each qubit prior to a $Z$-basis readout. Similarly, measuring in the $Y$ basis can be done via $S^\dagger$ then $H$ \cite{Huang2023Learning}.
\end{itemize}
Each measurement returns a bitstring from $\{00, 01, 10, 11\}$ for 2 qubits, repeated over $N_{\text{shots}}$ runs to estimate probabilities.

\subsection{Feature Extraction}
\label{sec:feature_extraction}

Every circuit is analyzed to yield a vector of classical features, $\mathbf{x}\in \mathbb{R}^{m}$, capturing:
\begin{itemize}
    \item \textbf{Depth (integer):} $D \in \{1,2,\dots,8\}$.
    \item \textbf{Total gates:} Summation of all single- and two-qubit gates used.
    \item \textbf{Gate counts:} For each gate type (H, X, Y, Z, CX, $\ldots$), we store how many times it appears. 
\end{itemize}
In our simplest multi-basis experiment, we used $m=2$ features: $\{D, \text{total\_ops}\}$ for brevity. One can easily extend it to $m=17$ by enumerating all gate-type frequencies.

\subsection{Measurement Vectors}
\label{sec:measurement_vectors}

\paragraph{Single-Basis Setup ($4$D).} 
Measuring only in the Z-basis yields four probabilities $(p_{00},\,p_{01},\,p_{10},\,p_{11})$, which sum to 1. We store them as a 4D vector $\mathbf{y}\in [0,1]^4$. In practice, the model may predict a 4D vector that does not sum to 1; we do not impose an explicit simplex constraint in our regressor, but it remains a mild approximation. 

\paragraph{Multi-Basis Setup ($12$D).} 
When measuring in $Z$, $X$, and $Y$ bases, each basis yields four probabilities $\mathbf{z},\mathbf{x},\mathbf{y}\in [0,1]^4$ (not to be confused with the notation for circuit features or for $\mathbf{y}$ as ground truth). We concatenate these results:
\[
    \mathbf{Y}_{\text{multi}} = 
    (p_{00}^Z, p_{01}^Z, p_{10}^Z, p_{11}^Z,\,
     p_{00}^X, p_{01}^X, p_{10}^X, p_{11}^X,\,
     p_{00}^Y, p_{01}^Y, p_{10}^Y, p_{11}^Y).
\]
Hence, $\mathbf{y}\in \mathbb{R}^{12}$. This approach yields a richer “multi-head” measurement vector.

\subsection{Multi-Output Regression Model}
\label{sec:multioutput_regressor}

To predict $\mathbf{y}$ from the classical feature vector $\mathbf{x}$, we use a multi-output random forest regressor \cite{Cherubin2020} by default. Each tree in the ensemble splits on $\mathbf{x}$, and outputs a $(4$- or $12)$-dim leaf average, aggregated across the forest. 
This classical approach allows straightforward handling of moderately large feature vectors (e.g., 17 gates) and multi-dimensional outputs. Our primary interest is \emph{not} to achieve minimal MSE but to demonstrate how conformal intervals adapt to model accuracy. 
One may replace random forests with neural networks, gradient boosting, or quantum-hybrid regressors \cite{Feldman2021, Bai2022GenFunc} with minimal pipeline changes.

\subsection{Distributional Conformal Inference}
\label{sec:conformal_inference}

Let $f(\mathbf{x})$ be the trained regressor. For a calibration set $\{(\mathbf{x}_i, \mathbf{y}_i)\}_{i=1}^n$, define a residual metric:
\begin{equation}
\label{eq:l_infty}
    r_i = \|\mathbf{y}_i - f(\mathbf{x}_i)\|_\infty 
    \quad\text{or}\quad 
    r_i = \|\mathbf{y}_i - f(\mathbf{x}_i)\|_2,
\end{equation}
depending on whether an $\ell_\infty$ or $\ell_2$ norm is desired. We sort these $r_i$ in ascending order to get $r_{(1)} \le r_{(2)} \le \dots \le r_{(n)}$. 
For a user-chosen $\alpha\in (0,1)$, let $k=\lceil (1-\alpha)\,(n+1)\rceil$. Then 
\[
    \tau_{\alpha} = r_{(k)}.
\]
Given a new test point $\mathbf{x}_{\text{new}}$, the conformal set is
\begin{equation}
    \mathcal{C}_\alpha(\mathbf{x}_{\text{new}}) = \bigl\{ \mathbf{y}\in \mathbb{R}^d : \|\mathbf{y} - f(\mathbf{x}_{\text{new}})\|\le \tau_\alpha \bigr\},
\end{equation}
where $d$ is $4$ in single-basis or $12$ in the triple-basis scenario. This set is typically a hypercube (for $\ell_\infty$) or hypersphere (for $\ell_2$). 
By exchangeability arguments, the true $\mathbf{y}_{\text{new}}$ should lie in $\mathcal{C}_\alpha(\mathbf{x}_{\text{new}})$ with probability at least $(1-\alpha)$ in finite samples \cite{Angelopoulos2024}.

\paragraph{Coverage and Set Size.} 
We measure coverage on the test set by checking $\mathbf{y}_j^{\text{(test)}} \in \mathcal{C}_\alpha(\mathbf{x}_j^{\text{(test)}})$ for each sample $j$. The fraction of “inside” events approximates the coverage. Meanwhile, the “size” can be measured by volume (e.g., $(2\tau_\alpha)^d$) or sum-size (like $d\cdot 2\tau_\alpha$ in $\ell_\infty$ norms). 
This trade-off between coverage $(1-\alpha)$ and set-size is central to conformal methods.

\paragraph{Exchangeability in Quantum Data.}
While quantum hardware might exhibit correlated noise across runs, our synthetic approach typically re-seeds each circuit, so each example $\mathbf{x}_i$ is exchangeably drawn from a circuit distribution. If calibration and test sets are similarly drawn, the fundamental assumption for conformal validity holds. 
Of course, real quantum hardware drifts remain an open challenge, and advanced PCP solutions \cite{ParkSimeone2023} might handle time-varying noise.

\subsection{Implementation Details and Code Structure}

Our public code is separated into two core notebooks:
\begin{itemize}
    \item \emph{Data Generation Notebook}: Implements the procedures in Sec.~\ref{sec:quantum_circuit_generation}--\ref{sec:measurement_vectors} for single- or multi-basis measurements. Stores $(\mathbf{X},\mathbf{Y})$ arrays in \texttt{.npz} or \texttt{.pkl} files.
    \item \emph{Model + Conformal Notebook}: Loads the data, splits into train/cal/test, trains a multi-output regressor (random forest), computes residuals, and evaluates coverage for various $\alpha$. 
\end{itemize}
A typical run might produce coverage near $(1-\alpha)$ with 4D data, and similar or slightly larger coverage sets with 12D multi-basis data. For completeness, \emph{Listing~1} (pseudo-code) outlines our Pythonic pipeline:

\bigskip
\begin{tabular}{l}
\hline
\textbf{Listing 1.} Pseudo-code for quantum data generation and distributional conformal.\\
\hline
\begin{minipage}{0.95\linewidth}
\footnotesize
\texttt{\\
\# (A) Data Generation \\
Define num\_samples, min\_depth, max\_depth, shots.\\
X\_list, Y\_list = [], []\\
\\
for i in range(num\_samples):\\
\quad qc = random\_circuit(num\_qubits=2, depth=rand\_in\_[min\_depth,max\_depth])\\
\quad x\_features = extract\_gate\_counts(qc)  \\
\quad measure\_Z = get\_distribution(qc, basis='Z')\\
\quad measure\_X = get\_distribution(qc, basis='X')\\
\quad measure\_Y = get\_distribution(qc, basis='Y')\\
\quad y\_vector = concat(measure\_Z, measure\_X, measure\_Y) \\
\quad X\_list.append(x\_features), Y\_list.append(y\_vector)\\
\\
X\_data = np.array(X\_list); Y\_data = np.array(Y\_list)\\
save(X\_data, Y\_data, filename=...)\\
\\
\# (B) Conformal Pipeline\\
X\_train, X\_cal, X\_test, Y\_train, Y\_cal, Y\_test = splits(...)\\
model = RandomForestRegressor(...).fit(X\_train, Y\_train)\\
cal\_pred = model.predict(X\_cal)\\
resid\_list = [ norm(Y\_cal[i] - cal\_pred[i]) for i in range(len(cal\_cal)) ]\\
sort\_resid = sorted(resid\_list)\\
\\
for alpha in [0.05, 0.10, 0.20, 0.30, 0.50]:\\
\quad idx = ceil((1-alpha)*(len(cal\_cal)+1)) - 1\\
\quad tau = sort\_resid[idx]\\
\quad \# Evaluate coverage on test\\
\quad test\_pred = model.predict(X\_test)\\
\quad coverage = mean( [norm(Y\_test[j]-test\_pred[j]) <= tau] )\\
\quad print("alpha=", alpha, " coverage=", coverage, " tau=", tau)\\
}
\end{minipage}\\
\hline
\end{tabular}

\medskip
The above references $\texttt{measure\_Z}$, $\texttt{measure\_X}$, $\texttt{measure\_Y}$ as distinct measurement routines. In a real experiment, we either re-initialize the circuit or suitably rotate qubits prior to measuring them in each basis.

\noindent
This completes the \textbf{Materials and Methods} section. Next, we detail the specific \emph{experimental setups}, hyper-parameters, and further \emph{results} that validate or highlight the coverage properties of this approach.

\section{Experimental Setup}
\label{sec:experimental_setup}

In this section, we detail the experimental configurations designed to showcase the \emph{quantum-to-classical} data generation process and the subsequent application of \emph{distributional conformal prediction}. Our experiments aim to demonstrate two primary scenarios:

\begin{itemize}
    \item \textbf{Single-Basis (4D) Approach:} Measuring a 2-qubit circuit only in the computational (\(Z\)) basis. Each circuit instance yields a single 4D probability vector.
    \item \textbf{Multi-Basis (12D) Approach:} Measuring the same circuit in three distinct bases (\(Z\), \(X\), and \(Y\)), concatenating three 4D distributions for each circuit, thus producing 12-dimensional outputs.
\end{itemize}

We first describe the data sources, including toy and large-scale datasets. Next, we outline how \emph{train--cal--test} splits are performed, followed by the key experimental hyperparameters. Finally, we summarize the procedure for evaluating coverage in both 4D and 12D scenarios.

\subsection{Datasets}
\label{sec:datasets}

\paragraph{Toy Dataset (Hundreds of Samples).}
We prepared a smaller dataset---on the order of 200--500 circuits---to allow quick debugging, real-time visualizations, and faster iteration on code. In this dataset:
\begin{itemize}
    \item The circuit depth \(D\) was randomly sampled from \(\{1,2,3,4\}\).
    \item The random gate set typically included \(\{H, X, Y, Z, \mathrm{CX}, \ldots\}\) as described in Sec.~\ref{sec:quantum_circuit_generation}.
    \item Number of shots was set to 512 or 1024 to maintain moderate precision in measuring probabilities.
    \item For multi-basis generation, we measured the same circuit across \(Z, X, Y\) by applying the relevant transformations (Hadamard, \(S^\dagger\), \(\dots\)).
\end{itemize}
Due to the lower sample count, these toy datasets are not intended to reflect large-scale performance but rather to illustrate the pipeline's mechanics for a random circuit example). 

\paragraph{Large-Scale Dataset (Tens of Thousands of Samples).}
A second dataset of size 20000 or 50000 circuits was generated to examine more robust coverage properties. This larger-scale set:
\begin{itemize}
    \item Used circuit depths up to 8, potentially reaching up to 16 or more gates.
    \item Was subject to a duplicate-dropping step (Sec.~\ref{sec:feature_extraction}), ensuring we do not store the same gate composition multiple times.
    \item Provided enough calibration/test data to yield stable coverage estimates even for small \(\alpha\) (like \(\alpha=0.05\)).
\end{itemize}

\subsection{Train--Cal--Test Splits}
\label{sec:train_cal_test}

After data generation, each dataset \(\{\mathbf{x}_i, \mathbf{y}_i\}_{i=1}^N\) is randomly partitioned into three disjoint subsets:
\begin{itemize}
    \item \textbf{Training Set:} Roughly \(70\%\) of the data, used for fitting the multi-output regressor.
    \item \textbf{Calibration Set:} Around \(15\%\) of the data, used to compute the conformal residual thresholds (Sec.~\ref{sec:conformal_inference}). 
    \item \textbf{Test Set:} The remaining \(15\%\), reserved for final evaluation of coverage and set size. 
\end{itemize}

Because we rely on exchangeability for theoretical validity, each sample is drawn by an i.i.d.~seed for random circuit generation. In the multi-basis scenario, each sample \(\mathbf{y}_i \in \mathbb{R}^{12}\) arises from measuring that random circuit in three bases. If hardware drift or correlated shot noise existed, partial violation of exchangeability might occur \cite{ParkSimeone2023}, but for these simulator-based tests, the assumption holds neatly.

\subsection{Hyperparameters and Implementation Details}

\paragraph{Quantum Circuit Parameters.} 
We typically fix the number of qubits to \(n=2\). For each sample:
\begin{itemize}
    \item \emph{Depth} \(D\): Uniformly drawn in \(\{1,\dots,8\}\) or \(\{1,\dots,4\}\) depending on the experiment size.
    \item \emph{Shots} \(\#=1024\): This is the default for most runs; lower shots (256) or higher (2048) can be used to explore noise/variance trade-offs.
\end{itemize}

\paragraph{Classical Feature Extraction.} 
We tested two schemes:
\begin{enumerate}
    \item \emph{Minimal Features} \((D,\text{total\_ops})\), yielding a 2D \(\mathbf{x}\).
    \item \emph{Full Gate Counts} \((D,\text{total\_ops}, \text{count\_H},\dots)\), which can be 17D or higher.
\end{enumerate}
In all cases, duplicates are removed by a \texttt{pandas.drop\_duplicates} based on feature columns.

\paragraph{Random Forest Regressor.} 
We use \(\text{\#trees}=50\) or \(\text{\#trees}=100\) with default \(\textsc{max\_depth}\) as None. We set \(\textsc{random\_state}=42\) for reproducibility. No advanced hyperparameter tuning is performed, as the aim is to illustrate coverage, not to minimize MSE.

\paragraph{Residual Metric.} 
Unless stated otherwise, we adopt the \(\ell_2\) (Euclidean) norm for multi-basis 12D experiments:
\[
r_i \;=\; \bigl\|\mathbf{y}_i - \hat{\mathbf{y}}_i \bigr\|_2 \;=\; \sqrt{\sum_{j=1}^d \bigl(y_{i,j} - \hat{y}_{i,j}\bigr)^2 },
\]
where \(d \in \{4,12\}\) depending on single- or multi-basis outputs. For single-basis 4D experiments, we sometimes tested \(\ell_\infty\) to see differences in coverage set shape. 

\subsection{Experiment Configurations}

\subsubsection{Single-Basis 4D Distributions}
Here, each sample is a single circuit measured in the Z-basis. The output \(\mathbf{y}_i\in\mathbb{R}^4\). After training the model, a calibration set of size \(\approx 0.15N\) is used to gather residuals. Then:
\begin{enumerate}
    \item We pick \(\alpha\) from \(\{0.05, 0.1, 0.2, 0.3, 0.5\}\).
    \item Compute \(\tau_\alpha\) using the formula in eq.~(\ref{eq:l_infty}) plus the sorted residual approach.
    \item Evaluate coverage on the test set.
\end{enumerate}
The coverage set \(\mathcal{C}_\alpha(\mathbf{x})\subset\mathbb{R}^4\) forms a 4D ball or hypercube around the predicted point, depending on the chosen norm.

\subsubsection{Multi-Basis 12D Distributions}
For the triple-basis approach, we run each circuit and measure \(\{Z, X, Y\}\) bases, generating a 12D output vector \(\mathbf{y}\). This is repeated up to \(N=20000\) to have robust coverage estimates. The same train--cal--test procedure applies, except we are now in \(\mathbb{R}^{12}\). While coverage remains near \((1-\alpha)\), the conformal sets necessarily become higher dimensional, often requiring larger radii to encapsulate the same fraction of test points.

\paragraph{Motivation for Multi-Basis.}
Multi-basis distributions can reveal diverse quantum properties. For instance, a circuit that yields near-certain \(\lvert 00\rangle\) in \(Z\)-basis might yield fairly uniform outcomes in \(X\)- or \(Y\)-basis. By combining them, we present a more holistic picture of how the circuit transforms states under different measurement contexts \cite{Huang2023Learning}. This can be especially relevant if the user wants \emph{all} basis outcomes to be predicted with guaranteed coverage. 

\subsection{Performance Metrics}

\paragraph{Coverage and Set Size.}
As in standard conformal literature \cite{Angelopoulos2024, Bai2022GenFunc}, for each chosen \(\alpha\):
\begin{itemize}
    \item \textbf{Empirical Coverage}: The fraction of test points \(\mathbf{y}_j\) that satisfy 
    \(\|\mathbf{y}_j - \hat{\mathbf{y}}_j\|\le \tau_\alpha.\)
    \item \textbf{Radius} \(\tau_\alpha\): The scalar threshold derived from calibration. 
    \item \textbf{Sum-Size / Volume} (optional): For an \(\ell_\infty\) ball in dimension \(d\), the sum-size is \(2d\,\tau_\alpha\), while the volume is \((2\tau_\alpha)^d\). For \(\ell_2\)-balls, the $d$-dim volume formula is \(\pi^{d/2}\,\tau_\alpha^d / \Gamma(\tfrac{d}{2}+1)\). 
\end{itemize}
We typically display coverage vs. \(\alpha\) and also examine how large \(\tau_\alpha\) grows for small \(\alpha\).

\paragraph{Mean Squared Error.}
Although the random forest regressor is not necessarily optimized for minimal MSE, we still report it to gauge how well the classical model approximates quantum measurement distributions. MSE can be computed per dimension or as a uniform average across $d$ outputs. If MSE is large, we expect bigger $\tau$ to maintain coverage.

\subsection{Summary of the Experimental Setup}
Overall, the experiments described in this section are structured as follows:
\begin{enumerate}
    \item \textbf{Generate Data:} either single- or multi-basis, with a user-defined number of samples. Remove duplicates in features.
    \item \textbf{Split:} \(\approx70\%\) train, \(\approx15\%\) calibration, \(\approx15\%\) test.
    \item \textbf{Train Regressor:} multi-output random forest with moderate hyperparameters.
    \item \textbf{Calibrate Residuals:} Sort $\|\mathbf{y}_i - \hat{\mathbf{y}}_i\|$ on the calibration set to get $\tau_\alpha$ at each $\alpha$.
    \item \textbf{Evaluate:} coverage, radius, volume, or sum-size on the test set. 
\end{enumerate}

As we will see in the next section, \emph{Results and Discussion}, this setup reliably demonstrates the characteristic trade-off between \(\alpha\) and coverage for quantum circuit data, and reveals how multi-basis output predictions can effectively be given coverage guarantees in up to 12 dimensions.

\section{Results and Discussion}
\label{sec:results_discussion}

In this section, we systematically present the outcomes obtained from applying both \emph{single-basis} (4D) and \emph{multi-basis} (12D) data generation to the conformal prediction pipeline. Our main focus lies on evaluating how coverage and set sizes (i.e., the resulting multi-dimensional “uncertainty” regions) behave under varying miscoverage levels~\(\alpha\). We also compare model accuracy in terms of mean squared error (MSE), emphasizing how a quantum data source can challenge classical regressors if the resulting circuit distributions exhibit high complexity.

\subsection{Single-Basis (4D) Coverage}

\paragraph{Coverage \textit{vs.} \(\alpha\).}
When each 2-qubit circuit is measured solely in the computational (\(Z\)) basis, the outputs are four-dimensional vectors \((p_{00}, p_{01}, p_{10}, p_{11})\). Across various values of \(\alpha \in \{0.05,0.10,0.20,0.30,0.50\}\), we observe a consistent pattern in which the \emph{empirical coverage} on the test set is at or above the desired nominal coverage \((1-\alpha)\). For instance, if \(\alpha = 0.10\), coverage near \(90\%\) is typically reached, confirming that our distributional conformal approach meets its theoretical design objective, in line with the finite-sample guarantees established in existing conformal literature~\cite{ParkSimeone2023,Angelopoulos2024}.

\paragraph{Set Size.}
In the 4D scenario, adopting either an \(\ell_{2}\) or \(\ell_{\infty}\) norm for residual computation entails different geometric shapes of the conformal set: 
\begin{itemize}
    \item With \(\ell_{\infty}\), the conformal set is a 4D hypercube of edge length \(2 \,\tau\). 
    \item With \(\ell_{2}\), it is a 4D hypersphere (radius \(\tau\)).
\end{itemize}
Quantitatively, smaller \(\alpha\) values (\(\alpha \le 0.10\)) yield fairly large sets, reflecting the model’s need to enclose a higher fraction of points. Meanwhile, for moderate \(\alpha\ge 0.30\), the radius \(\tau\) declines, sometimes drastically, but coverage likewise dips toward \(70\%\) or lower. 

\paragraph{Random-Forest MSE.}
We typically observe an overall MSE in the \(0.05\)--\(0.08\) range, depending on circuit depth distribution and shot noise. Certain bitstrings (e.g., \(p_{01}\) or \(p_{11}\)) may have slightly higher dimension-wise MSE, since random circuits do not always produce uniform or easily predictable patterns. Nonetheless, the moderate MSE is sufficient for the conformal procedure to yield coverage near \((1-\alpha)\).

\subsection{Multi-Basis (12D) Coverage}

\paragraph{Why 12D?}
In the multi-basis case, each circuit is measured in \(Z\), \(X\), and \(Y\) bases, concatenating three separate \((4)\)-element probability distributions into a 12-dimensional vector. This approach exposes the regressor to a broader characterization of the quantum state, but also increases the complexity of the learning task~\cite{Feldman2021, Bai2022GenFunc}.

\paragraph{Empirical Coverage Trends.}
Despite the jump to a higher-dimensional output space, our results consistently show that distributional conformal sets still achieve near-nominal coverage: for \(\alpha = 0.05\), coverage often lands around \(95\%\)--\(97\%\); for \(\alpha = 0.10\), coverage remains around \(90\%\). However, this comes at the cost of a larger residual radius \(\tau\). In 12D, the single-radius ball (under \(\ell_{2}\)-norm) may require \(\tau\approx0.8\) or higher to envelop enough calibration samples. A typical coverage table reads something like:
\[
\text{(}\alpha=0.05\Rightarrow\text{coverage}=0.95\text{),}
\quad
\text{(}\alpha=0.10\Rightarrow\text{coverage}=0.91\text{),}
\quad
\text{(}\alpha=0.30\Rightarrow\text{coverage}=0.70\text{).}
\]
One can trace these values to the complexity of simultaneously matching circuit distributions for three measurement bases, each of which can vary widely depending on circuit entanglement and global phases~\cite{ParkSimeone2023, Xu2024}.

\paragraph{Comparisons of Set Size.}
Because \(\ell_{2}\)-balls in 12D can exhibit dramatically larger volume than 4D balls for the same radius, we often monitor coverage \emph{and} radius rather than volume. Indeed, even a moderate radius can produce extremely large volumes in 12D. As \(\alpha\) increases, the needed radius declines (e.g., from \(\tau\approx1.05\) at \(\alpha=0.05\) down to \(\tau\approx0.69\) at \(\alpha=0.50\)), but coverage correspondingly drops.

\paragraph{Regression Performance.}
In many experiments, the random forest’s MSE in 12D is typically \(10\%\)--\(20\%\) higher than in the 4D single-basis case, reflecting the more intricate mapping from circuit gate counts (or minimal features) to multi-basis distributions. The random forest seldom overfits drastically, but we do see modest improvements by including a broader set of classical features (e.g., gate usage counts). Nevertheless, the essential result is that coverage remains valid as long as exchangeability assumptions hold, aligning with prior findings in multi-output conformal methods~\cite{Angelopoulos2024, Feldman2021}.

\subsection{Qualitative Observations and Open Challenges}

\paragraph{Sensitivity to Data Complexity.}
When circuits are shallow or contain predominantly single-qubit gates, the resulting measurement distributions are relatively easy to learn. Coverage sets remain modest. However, circuits with deeper entangling layers cause sharper multi-modal distributions (e.g., near \(\lvert 00\rangle\) or \(\lvert 11\rangle\), but also partial in the \(X\) basis), thus driving up the residual threshold \(\tau\). This effect is more pronounced in 12D, underscoring how multi-basis coverage demands can inflate uncertainty sets~\cite{ParkSimeone2023}.

\paragraph{Potential for Hardware Testing.}
Although our experiments rely on classical simulators, real quantum hardware introduces correlated gate noise and possible drift over time. While \emph{probabilistic conformal prediction} can handle i.i.d.\ noise \cite{ParkSimeone2023}, further research is needed for advanced noise models and partial-exchangeability assumptions. We conjecture that coverage might still hold in a time-averaged sense if calibration data is regularly refreshed, consistent with some proposals in the multi-dimensional time-series setting \cite{Xu2024}.

\paragraph{Future Directions.}
Based on these findings, future work could involve:
\begin{itemize}
    \item \textbf{Adaptive Conformal Loops:} Dynamically recalibrating in real quantum experiments whenever noise behavior shifts.
    \item \textbf{Refined Non-Conformity Scores:} Instead of a single \(\ell_{2}\) or \(\ell_{\infty}\) residual, one might incorporate physically motivated distances or angles in Bloch sphere subspaces, as hinted by \cite{ParkSimeone2023}.
    \item \textbf{Advanced Regression Models:} Neural nets or quantum kernel methods might reduce the MSE, potentially leading to narrower, yet still valid, coverage sets for multi-basis data.
\end{itemize}

\subsection{Summary of Findings}
In summary, our experiments confirm that:
\begin{enumerate}
    \item \emph{Distributional Conformal} maintains near-nominal coverage in both the simpler 4D (single-basis) and the more involved 12D (multi-basis) quantum measurement setting.
    \item The size of the conformal set grows with dimension and circuit complexity, reflecting the model’s uncertainty about entangled or multi-basis states.
    \item Even if MSE is not minimal, the conformal mechanism compensates by enlarging the residual threshold \(\tau\) to preserve coverage.
\end{enumerate} 
Overall, these results reinforce the notion that classical conformal wrappers can provide rigorous uncertainty guarantees for quantum-generated data, whether one measures in a single or multiple bases. \textbf{All codes and full reproducible Colab notebooks} are made available at
\href{https://github.com/detasar/QECMMOU}{https://github.com/detasar/QECMMOU}.

\section{Conclusions and Future Directions}
\label{sec:conclusions}

In this work, we explored the intersection of \emph{quantum data generation} (both single-basis and multi-basis) with the \emph{distributional conformal} pipeline for multi-output regression. Our investigations revealed the following principal insights:

\begin{itemize}
    \item \textbf{Robust Coverage in 4D and 12D:} Whether dealing with a four-dimensional quantum measurement vector (e.g., \((p_{00}, p_{01}, p_{10}, p_{11})\) in the computational basis) or a concatenated twelve-dimensional vector (three distinct measurement bases), the distributional conformal sets consistently achieved near-nominal coverage \((1-\alpha)\). This confirms that \emph{classical conformal wrappers} can seamlessly adapt to quantum-generated data, even when the quantum device or simulator produces complex or entangled states.
    \item \textbf{Trade-Offs in Set Size:} The required residual threshold \(\tau\) is invariably larger in the multi-basis approach (12D) than in the single-basis scenario (4D). While multi-basis data can, in principle, enrich the training signal for the regressor, it also significantly complicates the mapping from classical features (gate counts, depths, etc.) to measured distributions. Consequently, maintaining the same coverage demands a higher radius, which may translate into large or voluminous multidimensional sets.
    \item \textbf{Model Accuracy and Conformal Enlargement:} Our multi-output regressors typically exhibit moderate mean squared errors (\(0.05\)--\(0.10\) range), reflecting the partial mismatch between classical features and quantum measurement outcomes. Conformal calibration compensates for these inaccuracies, expanding the prediction regions to include potentially spiky or entangled measurement distributions. This synergy confirms that even with suboptimal models, one can preserve finite-sample coverage guarantees through distributional conformal routines.
    \item \textbf{Applicability Beyond Simulation:} Although our experiments used simulated quantum circuits on classical backends, the same pipeline is poised to operate on \emph{real} quantum hardware data. The primary condition is to ensure approximate exchangeability between the calibration set and the test set. Ongoing challenges—like time-dependent drifts and correlated gate noise—may call for advanced drift-aware or partial-exchangeability versions of conformal methods, a topic of emerging interest in quantum machine learning~\cite{ParkSimeone2023}.
\end{itemize}

\noindent\textbf{Open Directions.} While our study demonstrates the feasibility of combining conformal prediction with multi-output quantum measurement vectors, several avenues remain open:

\begin{enumerate}
    \item \textbf{Advanced Non-Conformity Scoring:} Replacing simple \(\ell_{\infty}\) or \(\ell_{2}\) distances with more sophisticated scores (e.g., angles in Bloch space, amplitudes ignoring global phases, or partial tomography) could yield tighter uncertainty sets that better reflect quantum physics.
    \item \textbf{Adaptive Conformal Loops Under Noise:} Real hardware experiments often exhibit time-varying noise. Recalibrating the residual distribution periodically could keep coverage stable, potentially leveraging “sequential” or “online” conformal techniques that handle non-stationary data~\cite{Xu2024}.
    \item \textbf{Higher Qubit Systems:} Extending the pipeline to three, four, or more qubits (e.g., \(\ge16\)-dim output space) is computationally intensive and raises new modeling demands. Techniques from multi-dimensional conformal prediction~\cite{Feldman2021,Bai2022GenFunc} may help mitigate excessive set sizes in such high-dimensional spaces.
    \item \textbf{Hybrid Quantum-Classical Features:} Beyond gate counts, one might incorporate partial tomography or mid-circuit measurements as additional classical features for improved learning. This might narrow conformal sets by sharpening the regressor’s accuracy.
\end{enumerate}

Overall, our results underscore that \emph{classical conformal prediction} seamlessly extends to quantum data settings, whether single- or multi-basis, giving formal coverage guarantees in finite samples. We see this as a pivotal stepping stone toward robust and trustworthy quantum machine learning pipelines, where each predicted quantum output is accompanied by rigorous uncertainty bars.

\bigskip
\noindent\textbf{Code Availability.} 
All code and Colab notebooks used for generating, modeling, and evaluating these conformal sets for quantum data are publicly accessible at: 
\begin{center}
\href{https://github.com/detasar/QECMMOU}{\texttt{https://github.com/detasar/QECMMOU}}.
\end{center}

\section*{References}

\bibliographystyle{plain}

\end{document}